# High-Performance, Hysteresis-Free Carbon Nanotube FETs via Directed Assembly

*Stephen A. McGill[1], Saleem G. Rao[1,†], Pradeep Manandhar[1], Seunghun Hong[2], and Peng Xiong[1,*]*

[1] Department of Physics & MARTECH, Florida State University, Tallahassee, Florida 32306, USA.

[2] Department of Physics & Nano-Systems Institute, Seoul National University, Seoul, Korea.

[*] Corresponding author.  Email: xiong@martech.fsu.edu

[†] Current address: Department of Physics, Western Illinois University, Macomb, IL 61455

**ABSTRACT** High-performance single-wall carbon nanotube field-effect transistors (SWNT-FETs) are fabricated using directed assembly and mass-produced carbon nanotubes (CNTs). These FETs exhibit operating characteristics comparable to state-of-the-art devices, and the process provides a route to large-scale functional CNT circuit assembly that circumvents problems inherent in processes relying on chemical vapor deposition (CVD).  Furthermore, the integration of hydrophobic self-assembled monolayers (SAMs) in the device structure eliminates the primary source of gating hysteresis in SWNT-FETs, which leads to hysteresis-free FET operation while exposing unmodified nanotube surfaces to ambient air.

Carbon nanotubes have remarkable physical and electrical properties that distinguish them as ideal components for nano- and molecular electronics[1].  Efforts devoted to incorporating them in device fabrication are probably best highlighted by progress in FETs[2-7], which are fundamental to many nanotube-based applications from computing to sensing.  So far, however, most high-performance

CNT-FETs have been fabricated using SWNTs grown on-substrate via CVD, while devices produced through 'directed assembly' and/or utilizing mass-produced have tried to keep pace. Nevertheless, CVD tends to produce large-diameter, small bandgap nanotubes, and the high-temperatures involved in the growth process often preclude integration with pre-existing circuit structures. For such reasons, high-performance FETs made by procedures more compatible with conventional semiconductor fabrication techniques are desirable and may represent a precursor to large-scale CNT circuit assembly for future industrial applications. Here, we demonstrate the directed-assembly[8,9] of SWNT-FETs using nanotubes grown by high-pressure decomposition of carbon monoxide (HiPCO), which produces large quantities of small-diameter, semiconducting SWNTs. These devices exhibit operating characteristics comparable with state-of-the-art FETs based on SWNTs from various sources, including CVD[10-12]. Notably, our devices can be operated *free of gating hysteresis while exposing clean SWNT surfaces to ambient air*. Previously, this hysteresis-free operation was possible only by encasing the FET in a polymer[13], rendering it difficult for further circuit integration and inoperable for certain applications, such as sensing. Furthermore, since our method does not require any high-temperature processing steps, it is compatible with conventional semiconductor device fabrication processes.

The SWNT-FETs are prepared by adapting a directed assembly technique[8] involving the controlled deposition of methyl-terminated (-$CH_3$) SAMs. These monolayers are placed either by microcontact printing or immersion to functionalize the metal electrodes and gate insulator, thereby creating a template for the subsequent placement and alignment of nanotubes. The schematic diagram depicting this procedure is shown in Figs. 1a-d. Five sets of source-drain electrodes each having a ~1-2 μm wide x 100 μm long gap are patterned via photolithography on degeneratively p-doped, 375 μm thick Si(100) substrates with 100 nm thermal oxides. The highly-doped Si eventually serves as a common back-gate for the FETs. Gold or palladium contacts (thickness ~ 20 nm) with an underlying Ti layer (thickness ~ 6 nm) are subsequently deposited via thermal evaporation. After metal deposition and before lift-off, an array of 4 μm wide 1-octadecanethiol (ODT, $CH_3[CH_2]_{17}SH$) lines each separated by 2 μm is printed onto the electrodes using a poly-dimethyl siloxane (PDMS) stamp coated with a 2 mM solution of ODT in ethanol (Fig. 1a). The ODT molecules bind to Au and Pd via their thiol functional groups[14,15], and the exposed methyl groups of the monolayer eventually direct the nanotubes to land on the remaining bare electrode surface and to *align* across the source-drain gap[8].

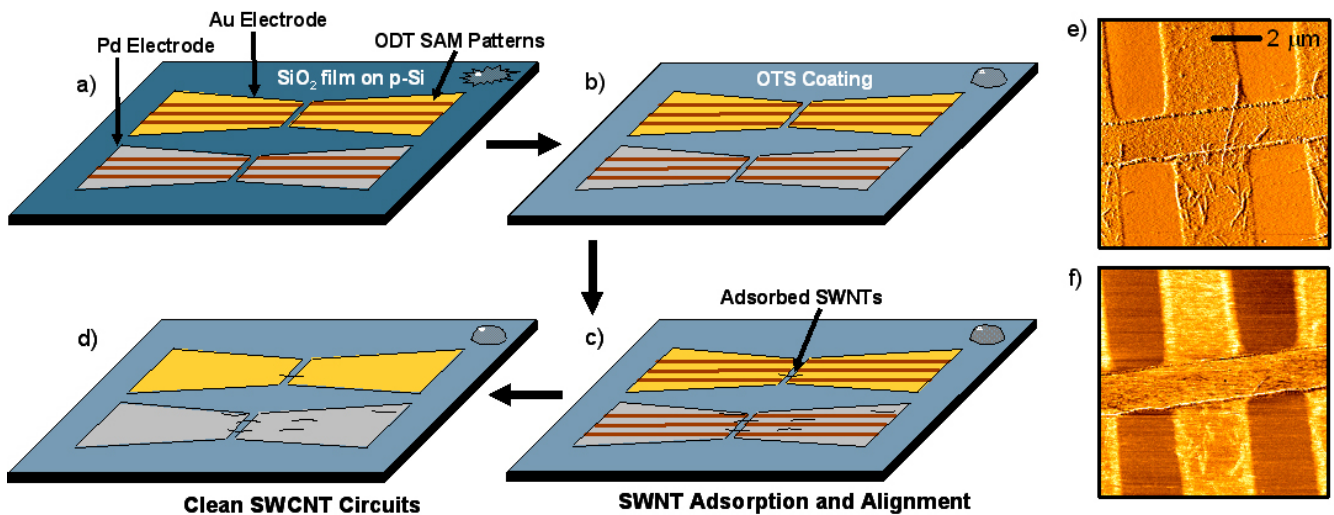

**Figure 1.** Assembly procedure and AFM images of a finished device. (a) Five sets of source-drain contacts (two shown) are fabricated via photolithography on a 1 cm x 1 cm chip. Lines of ODT are stamped onto the electrodes, shown after subsequent lift-off. (b) An OTS monolayer is deposited onto the $SiO_2$ surface making it hydrophobic. (c) SWNTs are deposited by immersion in a SWNT solution. (d) The substrate is heated up to 150-170° C to remove ODT layers from electrodes, resulting in clean SWNT circuits. (e) AFM image (10 μm x 10 μm) of an assembled device (unannealed) showing a nanotube bundle on Au electrodes bridging the source-drain gap. (f) Lateral force microscopy of the same view shows regions of reduced surface friction (darker contrast) revealing the stamped ODT pattern on the Au electrodes.

After lift-off and before deposition of nanotubes, an octadecyltrichlorosilane (OTS, $CH_3[CH_2]_{17}SiCl_3$) layer is grown on the $SiO_2$ via immersion in ~1mM solution in anhydrous hexane for ~12 hours (Fig. 1b). The OTS SAM is terminated with hydrophobic methyl groups and prevents nanotubes from collecting on the insulator and forming shorts between individual FETs. OTS also displaces water[16,17] from the surface of $SiO_2$ and plays an important role in removing gating hysteresis caused by water in contact with the CNT[13]. The substrates are then immersed in a solution of SWNTs in 1,2-dichlorobenzene (~0.03 mg/mL) for 1-3 minutes (Fig. 1c). The HiPCO SWNTs were obtained from Carbon Nanotechnologies Inc. and are ~1 nm in diameter (purity > 95 wt %). Also, all materials used in the deposition of the SAMs were used as received without further purification. Upon removal from solution, conducting paths between source and drain are established almost without failure, while isolation between individual devices remains unchanged (Figs. 1e,1f). Finally, the devices are annealed at 150-170° C in air for 12 hours. The annealing desorbs the ODT layers from the electrode surfaces[18], but OTS is not affected and $SiO_2$ remains hydrophobic (Fig. 1d). This fabrication process results in hysteresis-free SWNT devices with *clean, unmodified* CNT surfaces, which can be utilized for various applications such as sensing. Furthermore, since this is a scalable process based on only low-temperature processing steps, it can be an ideal method for producing integrated CNT circuits.

Some control over the number of assembled nanotubes can be achieved by varying the geometry of the PDMS stamp or the immersion time and concentration of the nanotube solution. Contact-mode AFM scans of the devices showed that up to five nanotubes or small bundles typically contacted both electrodes. Due to the random chirality of the nanotubes deposited from solution, these devices often exhibited transfer characteristics consistent with the presence of both metallic and semiconducting SWNTs. Figure 2 shows the gating behavior of typical devices as assembled on gold and palladium electrodes before annealing and the destruction of metallic SWNTs. The gating effect measured in each voltage loop was collected not with a continuous sweep but step-wise where ramping was paused to allow for settling and averaging in order to maximize the possibility of revealing any potential hysteresis. Devices fabricated with OTS passivation (step (b) in Fig. 1) do not show gating hysteresis (Fig. 2a), while those fabricated without OTS passivation have a large hysteresis (Fig. 2b) over the same voltage range. In general, this hysteresis has been attributed to the charging of traps which shield the nanotube from the gating field thereby shifting the "on/off" threshold of the device[19]. Water molecules on or near the nanotube have been identified as a primary source of these charge traps[13]. Our results clearly show that OTS passivation plays a central role in removing hysteresis. In prior fabrication methods, CNTs were in direct contact with hydrophilic $SiO_2$ surfaces or hydrophilic chemical functional groups such as amines (-NH$_2$). By assembling a SWNT on a hydrophobic SAM, hysteresis caused by water-based charge trapping is dramatically reduced.

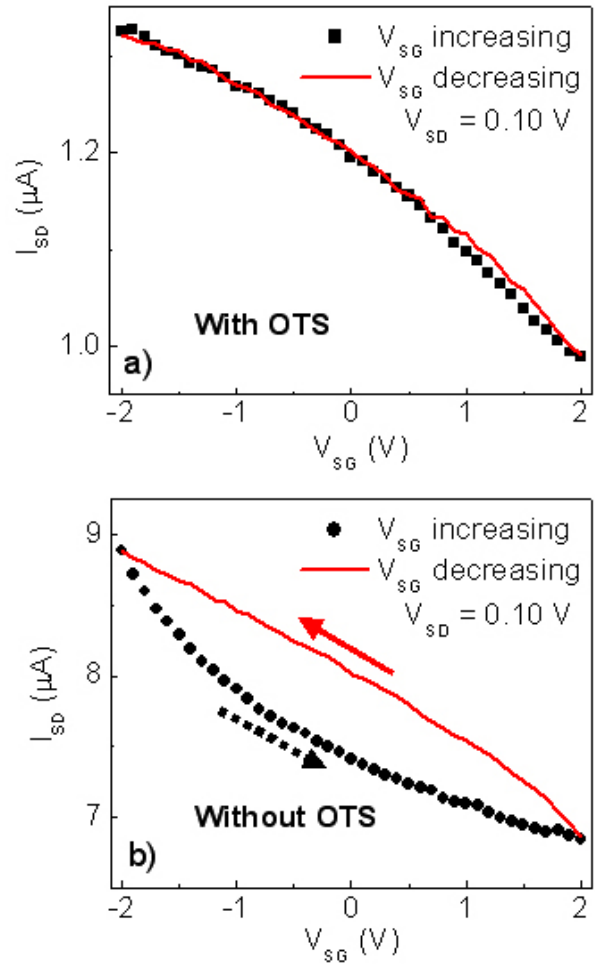

**Figure 2.** Gating effect in as-assembled devices before the selective destruction of metallic SWNTs. (a) Gating effect of an as-assembled CNT-FET with Au electrodes (unannealed) and OTS passivation of SiO$_2$ in an increasing (points) and reverse gating sweep (line). (b) Gating effect of an as-assembled CNT FET with Pd electrodes (unannealed) and without an OTS layer on SiO$_2$ in an increasing (points) and reverse gating sweep (line).

High-performance SWNT-FETs can be achieved by the selective destruction of metallic nanotubes[20]. Here, a large low-frequency (1-10 Hz) source-drain bias is applied to break metallic tubes while applying a positive gating voltage to place semiconducting tubes in the "off" state. Afterwards, the "on/off" ratio of the FET can increase up to $10^6$, and high-performance FET operation comprised of solely semiconducting SWNTs is realized. The progression of an as-assembled device to an all-semiconducting FET is shown in Fig. 3 (inset). Detailed characterization of hysteresis in FETs has been performed after the destruction of metallic nanotubes (Fig. 3). Generally, as in Fig. 3, gating voltages within ±2 V always produced hysteresis-free transfer characteristics for source-drain biases up to ±200 mV. We emphasize that *these voltage ranges are sufficient for full FET operation*. We have found that hysteresis can reappear in a larger gating loop (Fig. 4a). Two well-discussed origins of hysteresis for SWNT-FETs are: 1) adsorbed water[13] and 2) charge traps inside the gate dielectric[19]. The hysteresis appearing in the extended ±4 V gating loop (Fig. 4a) could not be reduced by placing the FET in a vacuum (3 x $10^{-6}$ torr) for more than 12 hours, which implies that the origin of the remaining hysteresis is not due to water adsorbed on the SWNT.

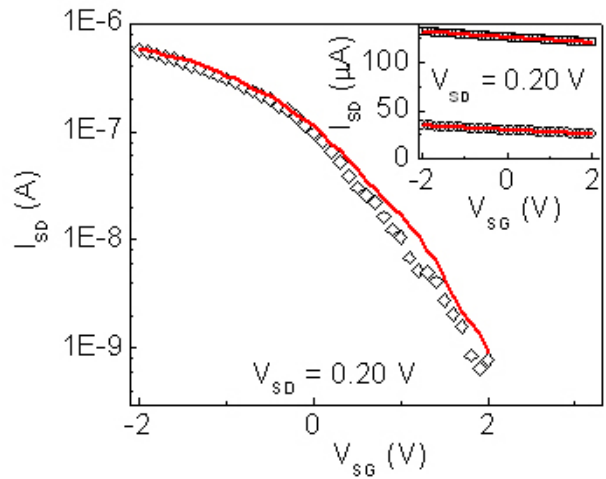

**Figure 3.** Creation of an all-semiconducting SWNT-FET via the selective destruction of metallic SWNTs. Transfer characteristic of an all-semiconducting device obtained by selectively destroying metallic nanotubes in an increasing (points) and reverse gating sweep (line). Inset shows gating effect of the device as-assembled and intermediate to the final FET stage in increasing (points) and reverse sweeps (lines).

Through our ability to eliminate gating hysteresis from SWNT-FETs, we have been able to gain a clearer understanding of its possible causes. By comparing the transfer characteristics of the same FET operated in the hysteresis-free range and then in the extended gating range as in Fig. 4a, one can see that the opening of a hysteresis loop corresponds primarily to a positive shift of the decreasing gating field sweep (+4 V to -4V). This shift of the threshold voltage is consistent with the direction expected if the hysteresis is due to trapping of electrons. In contrast, when the same comparison is made between a FET with an OTS monolayer and one without it, a markedly different result is obtained. In this case (Fig. 4b) the opening of a hysteresis loop in the device without OTS primarily corresponds to a negative shift of the increasing gating field sweep (-4 V to +4 V). The direction of this shift that occurs when the nanotube is in a hydrophilic environment is almost exactly opposite to that observed in Fig. 4a. A similar type of shift was previously reported in the transfer characteristics of other nanotube FETs and attributed to the gate-induced motion of ionic species mobile in water[21]. The contrasting hysteretic behaviors shown in Figs. 4a and 4b provide convincing evidence that these threshold shifts stem from two separate sources.

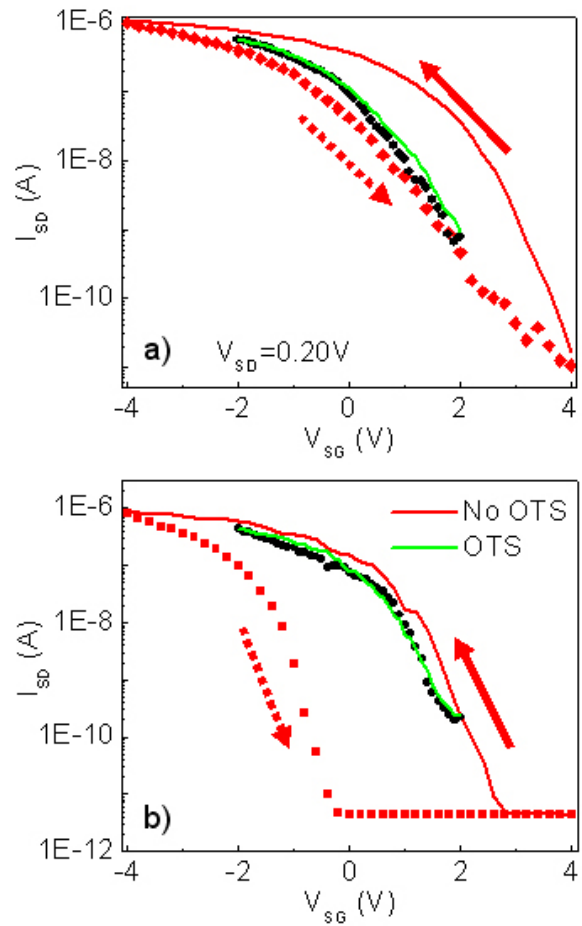

**Figure 4.** Oppositely directed threshold shifts caused by different hysteresis sources. (a) Transfer characteristics of the device from Fig. 3 in a wider gating loop compared with its hysteresis-free operation. (b) Transfer characteristics of a separately fabricated hysteresis-free FET compared with that of a device without an OTS layer. $V_{SD}$ for the hysteresis-free device was -0.1 V ($|I_{SD}|$ shown) and 1 V for the FET without OTS.

The hysteresis appearing after a large gating bias is applied may be linked to the quality of the substrate oxide used for device fabrication. Hysteresis in Si MOSFETs is due to interface and bulk oxide trap charges that can change state with gate bias[22]. Injection into bulk oxide traps can begin when electric fields of ~3 x $10^5$ V/cm are developed in the dielectric[22]. The data in Fig. 4a can be used to assess whether such traps can plausibly account for the additional hysteresis observed therein. The capacitance per unit length between a nanotube and the gate can be approximated[23] as $C/L \approx 2\pi\kappa\varepsilon_0/\ln(2t/r)$ in which $\kappa = 2.5$ is the average dielectric constant[23], t = 100nm is the dielectric thickness, and r = 0.5 nm is the nanotube radius. With this equation, one obtains $C/L \approx 23$ aF/μm = 150e$^-$/μmV, where e$^-$ is the electron charge. Using the voltage shift, $\Delta V \approx 2$ V in Fig. 4a, one can estimate the number of trapped charges for a 1 μm long device as $C\Delta V \approx 300$e$^-$. Supposing that each charge corresponds to a single trap yields a defect density of ~5 x $10^{10}$ cm$^{-2}$, which is certainly within range[19,22] for device grade oxides. It is notable that fields of ~3 x $10^5$ V/cm leading to charge trapping can be achieved in this approximation with as little as 0.1 V gating bias.

Although gating hysteresis may be useful in some novel device applications such as CNT-based memory cells[19], it has posed a persistent challenge in most electronic applications. In our method, an OTS coating on SiO$_2$ effectively eliminates the primary hysteresis due to water directly adsorbed on the SWNTs (Fig 3). The OTS layer is known to remove a thin film of water bound to silanols (SiOH) on the SiO$_2$ surface and, more importantly, appears to prevent any significant adsorption of water to the SWNTs though they are exposed to air. The hysteresis in wide gating loops may make hysteresis-free operation difficult for ambipolar FETs operating near the inversion regime[24], but this should be readily solved using different dielectrics[25]. In other approaches, this (primary) hysteresis has been addressed by spinning a thick layer of polymethyl methacrylate (PMMA) over the entire device[13]. This technique also reduces hysteresis but makes large-scale circuit integration more difficult and renders the nanotubes inoperable as sensors. Since our method does not use any coatings over SWNTs and exposes clean nanotube surfaces, the fabricated FET can be readily utilized for various sensor applications[26].

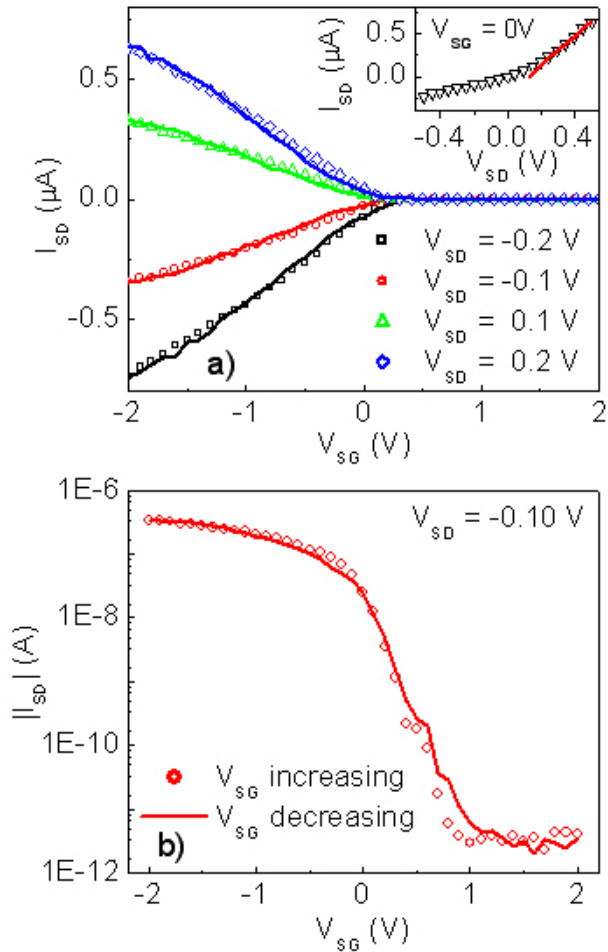

**Figure 5.** SWNT-FET operation after long-term storage. (a) Transfer characteristics of an all-semiconducting FET with Pd electrodes two months after its fabrication in increasing (points) and reverse gating sweeps (lines) for a variety of source-drain biases. (Inset) I-V characteristic for the same device at $V_{SG} = 0$ V. The line shows the result of fitting the forward-biased data (nablas). (b) A closer inspection of the data for $V_{SD} = -0.10$ V on a log scale (y-axis) confirms hysteresis-free operation was preserved throughout storage.

The performance characteristics of these FETs as well as the permanency of the removal of hysteresis are demonstrated through measurements on a Pd-electrode device taken over two months after its initial fabrication (Fig. 5a). As shown in Fig. 5b, these FETs can be comparable to state-of-the-art FETs and represent an important advance for those based on HiPCO-grown SWNTs and/or fabricated via directed-assembly methods[10,27]. This device demonstrated an "on/off" ratio of nearly $10^6$ with an "on" current approaching 1 µA. The inverse subthreshold slope, $S^{-1} = dV_{SG}/d(\log I_d)$, was 230 mV/dec (with 100 nm thick gate oxide) at room temperature. As shown in Fig. 5a (inset), the forward-bias region of the zero-gate I-V extrapolated back to an intercept of 130 meV to give a rough lower bound to the Schottky barrier height[28] at the Pd-nanotube interface. Operation nearly free from Schottky barriers has been demonstrated in devices with large (~3 nm) diameter SWNTs grown by CVD on Pd contacts[11]. For our SWNTs with smaller diameters and therefore larger bandgaps, Schottky barriers remained. Improved selectivity of nanotubes and redesigned electrode structure can lead to gains in FET performance by controlling the "on/off" threshold and decreasing $S^{-1}$. For example, $S^{-1}$ in Schottky-barrier FETs varies with the type and thickness of the dielectric used as well as the electrode geometry[29-31]. The same parameters can also affect the coupling of the nanotube to the gate and its effective doping level[11]. The method of assembly used here does little to restrict these degrees of freedom and so leaves room for further improvement in SWNT-FET operation.

While various self-assembly procedures to fabricate nanotube circuits have been reported, our methods and results differ significantly from these prior studies. In several cases, nanotubes were first deposited on oxides functionalized with a polar-terminated SAM and then source-drain electrodes were deposited afterwards[10,27,32]. In our case, nanotubes are the last elements to be deposited which minimizes the opportunity for contamination from subsequent processing steps. In another technique, end-functionalized nanotubes were assembled onto pre-defined contacts[33]. Our method uses unmodified nanotubes to eliminate concerns over affecting their electrical properties or needing to subsequently remove the attached groups. Finally, reduction of hysteresis has been demonstrated by depositing PMMA[13] and by rigorous cleaning of the substrate[21]. In the former case, PMMA covers the entire device highly limiting its use for various applications. Rigorous cleaning, as in the latter case, also reduces hysteresis but provides no protection from future contamination.

In summary, we demonstrated the fabrication of high-performance, hysteresis-free SWNT-FETs using HiPCO-grown nanotubes and a directed-assembly method. The OTS layer on the dielectric effectively removes the gating hysteresis caused by adsorbed water, while leaving clean CNT surfaces. Even though one still needs a high-yield method to separate semiconducting and metallic SWNTs for industrial-level production of CNT-based integrated circuits, we clearly demonstrated that the directed-assembly process can be utilized to assemble HiPCO-grown nanotubes to fabricate high-performance SWNT-FETs comparable with the state-of-the-art FETs made with a variety of strategies. Since our method can be expanded for large scale assembly of CNT-FETs without any high-temperature processing, it can be an ideal method for the industrial-level fabrication of CNT-based electronic devices in the future. The strategy can be readily adapted for the bottom-up assembly of similar devices based on semiconductor nanowires.

This work was supported by NSF NIRT ECS-0210332 and FSURF PEG. SH acknowledges financial support from the KOSEF through NRL and partial support from MOCIE.

1. McEuen, P.L. *Physics World* **2000**, *13*, 31-36.

2. Avouris, P. *MRS Bulletin* **2004**, *29*, 403-410.

3. Javey, A.; Wang, Q.; Kim,W.; Dai, H. *IEDM Tech. Digest* **2003**, 31-34.

4. McEuen, P.L.; Fuhrer, M.S.; Park, H. *IEEE Trans. Nanotechnol.* **2002**, 78-85.

5. Bachtold, A.; Hadley, P.; Nakanishi, T.; Dekker, C. *Science* **2001**, *294*, 1317-1320.

6. Bockrath, M.; Liang, W.; Bozovic, D.; Hafner, J.H.; Lieber, C.M.; Tinkham, M.; Park, H. *Science* **2001**, *291*, 283-285.

7. Rueckes, T.; Kim, K.; Joselevich, E.; Tseng, G.Y.; Cheung, C.-L.; Lieber, C.M. *Science* **2000**, *289*, 94-97.

8. Rao, S.G.; Huang, L.; Setyawan, W.; Hong, S. *Nature* **2003**, *425*, 36-37.

9. Liu, J.; Casavant, J.; Cox, M.; Walters, D.A.; Boul, P.; Lu, W.; Rimberg, A.J.; Smith, K.A.; Colbert, D.T.; Smalley, R.E. *Chem. Phys. Lett.* **1999**, *303*, 125-129.

10. Johnston, D.E.; Islam, M.F.; Yodh, A.G.; Johnson, A.T. *Nature Materials* **2005**, *4*, 589-592.

11. Javey, A.; Guo J.; Wang, Q.; Lundstrom M.; Dai, H. *Nature* **2003**, *424*, 654-657.

12. Chen, J.; Klinke, C.; Afzali, A.; Avouris, P. *Appl. Phys. Lett.* **2005**, *86*, 123108-123110.

13. Kim W.; Javey, A.; Vermesh, O.; Wang, Q.; Li, Y.; Dai, H. *Nano Letters* **2003**, *3*, 193.

14. Nuzzo, R.G.; Dubois, L.H.; Allara, D.L. *J. Am. Chem. Soc.* **1990**, *112*, 558-569.

15. Love, J.C.; Wolfe, D.B.; Haasch, R.; Chabinyc, M.L.; Paul, K.E.; Whitesides, G.M.; Nuzzo, R.G. *J. Am. Chem. Soc.* **2003**, *125*, 2597-2609.

16. Wang, Y.; Lieberman, M. *Langmuir* **2003**, *19*, 1159-1167.

17. Wang, M.; Liechti, K.M.; Wang, Q.; White, J.M. *Langmuir* **2005**, *21*, 1848-1857.

18. Nuzzo, R.G.; Zegarski, B.R.; Dubois, L.H. *J. Am. Chem. Soc.* **1987**, *109*, 733-740.

19. Radosavljević, M.; Freitag, M.; Thadani, K.V.; Johnson, A.T. *Nano Letters* **2002**, *2*, 761-764.

20. Collins, P.G.; Arnold, M.S.; Avouris, P. *Science* **2001**, *292*, 706-709.

21. Bradley, K.; Cumings, J.; Star, A.; Gabriel, J.-C.P.; Grüner, G. *Nano Letters* **2003**, *3*, 639.

22. Wolf, S.; Tauber, R.N. *Silicon Processing for the VLSI Era*, (Lattice Press, Sunset Beach, CA, 1986).

23. Martel, R.; Schmidt, T.; Shea, H.R.; Hertel, T.; Avouris, Ph. *Appl. Phys. Lett.* **1998**, *73*, 2447-2449.

24. Martel, R.; Derycke, V.; Lavoie, C.; Appenzeller, J.; Chan, K.K.; Tersoff, J.; Avouris, Ph. *Phys. Rev. Lett.* **2001**, *87*, 256805-256808.

25. Ye, P.D.; Wilk, G.D.; Yang, B.; Kwo, J.; Chu, S.N.G.; Nakahara, S.; Gossmann, H.-J.L.; Mannaerts, J.P.; Hong, M.; Ng, K.K.; Bude, J. *Appl. Phys. Lett.* **2003**, *83*, 180-182.

26. Staii, C.; Chen, M.; Gelperin, A.; Johnson, A.T. *Nano Letters* **2005**, *5*, 1774-1778.

27. Plank, N.O.V.; Ishida, M.; Cheung, R. *J. Vac. Sci. Technol. B* **2005**, *23*, 3178-3181.

28. Fuhrer, M.S.; Nygård, J.; Shih, L.; Forero, M.; Yoon, Y.-G.; Mazzoni, M.S.C.; Choi, H.J.; Ihm, J.; Louie, S.G.; Zettl, A.; McEuen, P.L. *Science* **2000,** *288*, 494-497.

29. Heinze, S.; Tersoff, J.; Martel, R.; Derycke, V.; Appenzeller, J.; Avouris, Ph. *Phys. Rev. Lett.* **2002**, *89*, 106801-106804.

30. Appenzeller, J.; Knoch, J.; Derycke, V.; Martel, R.; Wind, S.; Avouris, Ph. *Phys. Rev. Lett.* **2002**, *89*, 126801-126804.

31. Heinze, S.; Tersoff, J.; Avouris, P. *Appl. Phys. Lett.* **2003**, *83*, 5038-5040.

32. Auvray, S., Derycke, V., Goffman, M., Filoramo, A., Jost, O., and Bourgoin, J.-P. *Nano Letters* **2005**, *5*, 451-455.

33. Li, X., Liu, Y., Shi, D., Sun, Y., Yu, G., Zhu, D., Liu, H., Liu, X., and Wu, D. *Appl. Phys. Lett.* **2005**, *87*, 243102-243104.